# Reducing research bureaucracy in UK higher education: Can generative AI assist with the internal evaluation of quality?

*Gordon Fletcher, Salford Business School, University of Salford*

*Saomai Vu Khan, Salford Business School, University of Salford*

*Aldus Greenhill Fletcher, Independent Scholar*

g.fletcher@salford.ac.uk

*Abstract:*

This paper examines the potential for generative artificial intelligence (GenAI) to assist with internal review processes for research quality evaluations in UK higher education and particularly in preparation for the Research Excellence Framework (REF). Using the lens of function substitution in the Viable Systems Model, we present an experimental methodology using ChatGPT to score and rank business and management papers from REF 2021 submissions, "reverse engineering" the assessment by comparing AI-generated scores with known institutional results. Through rigorous testing of 822 papers across 11 institutions, we established scoring boundaries that aligned with reported REF outcomes: 49% between 1* and 2*, 59% between 2* and 3*, and 69% between 3* and 4*. The results demonstrate that AI can provide consistent evaluations that help identify borderline evaluation cases requiring additional human scrutiny while reducing the substantial resource burden of traditional internal review processes.  We argue for application through a nuanced hybrid approach that maintains academic integrity while addressing the multi-million pound costs associated with research evaluation bureaucracy. While acknowledging these limitations including potential AI biases, the research presents a promising framework for more efficient, consistent evaluations that could transform current approaches to research assessment.

## 1.     Introduction: The pragmatics of REF bureaucracy

The assessment of research quality is a critical component of UK academia for determining funding allocation (Thelwall *et al.*, 2023), influencing institutional reputation (Blackburn et al. 2024) and shaping individual career progression (Wróblewska, 2024). In the United Kingdom (UK), the Research Excellence Framework (REF) sets out the principles and requirements that define the quality of research outputs generated during each seven-year cycle with a clear incentive for institutional participation being the allocation of significant Quality-related Research (QR) funding from the UK government. With the incentive for institutions to participate clearly established, the process of internal peer review, used by the majority of institutions to determine the final pool of papers for each subject-based Unit of Assessment, is resource-intensive. All forms of peer review are a fundamental tenet of academic work that inspire action, self-reflection or inform a reviewer. However specific forms of review can also be regarded as a burden that can distract from the development of new knowledge and progressing one's own career (Ferraris, 2017; Babin and Moulard, 2018; Severin and Chatway, 2021). Nonetheless, there is an imperative to create the best possible pool of outputs for submission and this requires robust and accurate processes. For the purpose of the REF process, reviewing does not require a rank ordered list but it does require sufficient accuracy to determine what papers fall inside the submission pool with those papers sitting on this margin being the most challenging to accurately determine for inclusion or exclusion. This research explores the role of generative artificial intelligence (GenAI) in a supporting role to the existing internal processes that allows greater focus to be given to those works that do sit on the boundary of inclusion while simultaneously reducing overall resource burden. We consider the efficiency aspects of this technology-assisted approach as well as the benefits of improved consistency, comparability and precision in the overall evaluation of research outputs. To better understand the organisational implications of adopting GenAI for internal evaluation, we draw upon Beer's Viable System Model (VSM) (Beer, 1981) as a framework for theorising the role of AI-assistance as a viable functional substitute within higher education institutional architecture. This theorisation positions GenAI in the context of the VSM as a means of bringing stability to a system that is under threat from competing external pressures and the necessity of completing this repetitive and labour intensive work. While the exploration of VSM is directly applied to the use of GenAI for reviewing existing outputs, the theoretical potential for positioning the use of AI across an organisation in this manner is far more expansive and offers the potential to guide decision making regarding the deployment of this technology in a useful and productive manner. This is all the more important as AI-based technology can now be regarded as a commonplace feature of organisations, with Roberts and Candi (2024) observing that the use of AI in innovation processes is already "high and widespread," with over half of the firms they surveyed using AI in their projects.

*1.1 Why do internal peer review?*
Our position is that peer reviewing of all types are productive activities that ensure academia remains collectively relevant and responsive (Bedeian, 2004). Review is the basis for all quality control in academia but should also be a mutually beneficial process. The specific task of conducting internal reviews for the purpose of external national research assessment is to make a judgement regarding the relative merits of eligible outputs. This additional review cycle exists solely to decide the pool of already published works that will be submitted. And it is one reason that some UK research managers label it internal evaluation - rather than peer review - and to describe the external process as an assessment (Curry *et al.*, 2022). During the formal process of REF assessment, a third review by an expert who has no affiliation to the author or their institution is undertaken confidentially in order to assign a score ranging from Unclassified and 1* through to the highest, 4*. The percentage of outputs assigned each of these scores for each institutional submission to a Unit of Assessment is then reported as one key output of the REF process. These steps are important as the assessment outcomes determine the government distribution of recurring funding for institutional research. In England alone, this currently equates to around £2 billion annually (UKRI, 2023). And the assessment has direct financial consequences, as this QR funding to institutions is only distributed to the proportion of papers with a 3* or 4* score.

Subsequently, third parties, including national UK newspapers, use these percentages to calculate GPA scores for institutions and their submitted subject areas as well as league table calculations. These additional calculations influence institution choice among prospective students and are a further indirect financial incentive to engage the assessment with the best possible submission pool of works.

These are all clear reasons for engaging in the external assessment with a strong process as institutions want to make the best possible decisions regarding the works that are submitted. However, recognising all of these points does not automatically lead to the conclusion that peer evaluation is the only way of generating insights for the decision making process. Some institutions minimise this burden by reverting to journal rankings, citation counts or other combinations of metrics. But as the Declaration of Research Assessment (DORA) says, "Journal Impact Factors can be manipulated (or "gamed") by editorial policy and data used to calculate the Journal Impact Factors are neither transparent nor openly available to the public" (DORA, n.d.). This mitigates against the use of solely journal-based metrics to evaluate individual outputs. Invariably "good" papers in "bad" journals might be overlooked and "bad" papers in "good" journals will be celebrated well beyond their inherent merits. The core premise of DORA says that "it is thus imperative that scientific output is measured accurately and evaluated wisely." But this statement does not determine that peer evaluation is the exclusive approach that can be used in selecting papers for national research assessment programmes. This observation is the core premise for exploring the potential for alternative, more technology-based solutions that can address the same challenge and productively assist reviewers.

*1.2 The Burden of Internal Review*
Peer review is a demanding process that requires significant time and effort from academics (Aczel *et al.*, 2021). Each additional step involving review, evaluation or assessment multiplies the time commitment, diverting resources from other essential activities, including the process of research itself. Moreover, the management of the peer review process, such as anonymisation and score calculation, adds to the overall workload. The scale of these activities for the UK is reflected in the Cost Evaluation Report (Neto *et al.*, 2023) which defines reviewing and negotiating outputs as the second most time-consuming aspect of preparation within Units of Assessment (UoA), accounting for 29% of the time spent. The total cost of REF preparation at this level is estimated to be £284 million, with 92% of these costs are attributed to UoA review groups, champions and identifying eligible staff. Across the broader process, including UoA and institutional activities, the single largest cost element were review groups, at an estimated 34% of the £430 million spent (Neto *et al.*, 2023: 5, 12, 30). The potential for alternative methods to mitigate this burden would have a transformative effect on the broader research environment, allowing for substantial time and resources to be released back into generating impactful research activities and outputs. Just (2024) portrays the role of AI in organisations as a "non-human innovation intermediary" that can take on many early-stage tasks including "forecasting trends, illustrating ... landscapes, filtering out distinctive contributions [and] matching problems with solutions." This extends consideration of AI's role to the many functions of any organisation that are currently laboriously done through committee actions or repeated operations undertaken by multiple individuals to ensure consistency that can now be completely automated,

*1.3 The Promise of Generative AI*

In October 2022, the *Future Research Assessment Programme* (FRAP) (JISC, 2022) reported on the potential to utilise AI within assessment (Thelwall *et al*., 2022). This report is noteworthy for representing the moment before the release of OpenAI's first public generative AI interface the following month. Nonetheless, the 2022 report indicates positive results and speculates on even greater potential if an AI model were to be trained on a significant corpus of disciplinary work. While OpenAI and others technology that has emerged do not fully disclose the source of their training data there is strong indications that at least part of the available academic corpus has already been ingested into these models' training data. This raises new opportunities with each new model and this potential has been explored continuously since 2021 (Checco *et al*., 2021 Thelwall *et al*., 2023a; Mehta *et al*., 2024; Thelwall and Yaghi, 2024; Thelwall, 2025). However, the prospect of using AI to independently internally evaluate a paper's quality in REF terms - assessing its rigour, originality, and significance - remains largely unexplored with the exception of Kousha and Thelwall (2023) and Thelwall *et al*'s work (2022, 2023a, 2024, 2025).

There is a general reluctance to use AI for general pre-publication journal reviewing over concerns about the technology's ability to provide accurate evaluations, as well as potential ethical considerations regarding the integrity of the review (Leung *et al*., 2023; Singh, 2023; Chawla, 2024; Cheng *et al*., 2024; Liang et al., 2024; Lindebaum and Fleming, 2024; Tyser *et al*., 2024). The sentiment is founded on the principle that peer assessment is "the gold standard for ensuring the quality and credibility of research publications" (Bhosale and Kapadia, 2023), leading Mehta et al. (2024: 2) to caution that tools like ChatGPT, "should not be relied on as the sole means of review." Among the ethical concerns is the expectation that "[r]eviewers are expected and trusted to uphold confidentiality throughout the review process" (Bhosale & Kapadia, 2023). The use of generative AI and LLMs may violate this confidentiality during a pre-publication process, as the technology could then process this data to generate new content (Ramstad & Sanders, 2023). There is a potential for paper manuscripts that are reviewed through public interface generative AI tools to become absorbed into the overall training data prior to the publication of the paper itself. Although the statistical likelihood of AI reassembling an unpublished paper's conclusions is effectively zero it is still a prospect that makes authors uneasy. Fortunately, the post-publication evaluation assessment of research - especially with open publishing practices - does not raise these concerns.

Thelwall (2025) also warns that replacing human reviewers entirely may lead to authors learning how to "game the system" by writing with intent to return high scores from the AI, rather than to advance knowledge with future readers in mind, which could "degrade the core informational functions of research articles". This caution does imply that current systems of national research evaluation are not in some way already "gamed" to reach the most favourable outcomes. More critically, hesitations regarding the gaming of a technology-based system that is documented and transparent insofar as it is an information system is of a different order of concern to the embedded networks of social capital that shapes the unconscious bias found in entirely human-based selection and reviewing. Generative AI and LLMs do have acknowledged and systematic biases (Wei *et al.*, 2025) that consistently pervade the entire system, but human-based systems are erratically influenced by who knows who, co-authorships, which institution are associated with an output, its publisher and a lengthier list of variables that unconsciously influence human decision-making that is further exacerbated by time pressures that remove the opportunity for reflection (Checco *et al.*, 2021: 2). AI may be imperfect but it is consistently imperfect in a way that having many humans within a process cannot be.

Nonetheless, given these stated concerns, it is important to distinguish between review for publication and review for internal evaluation. Our proposition and experimental method uses generative AI to assess the quality of papers for a specific purpose, focusing on internal review towards a national external research assessment. As a result, the scores provided by AI are not about influencing decisions regarding a paper's credibility for publication and, as a result, preserves the integrity of the journal peer review process. As the papers in question are already publicly available through institutional and publisher repositories, the process we document does not breach confidentiality or any rights ownership. Going further, our use of the OpenAI applications programmer interface (API) for the experiment does not unwittingly submit papers into the training data of the LLM.

We argue that generative AI presents a promising assistant to human-based internal evaluation. By leveraging the extensive and improving knowledge and processing capabilities of LLMs, institutions could potentially automate much of the evaluation of outputs, reducing both the time and people time needed for this process (Chiarello et al., 2024) that collectively represents a multi-million pound saving against constrained budgets. The statistical nature of large language models also suggests they can deliver consistent and precise evaluations, potentially surpassing human reviewers in terms of comparability, accuracy and efficiency. In many respects the opportunity is transformative for the consistency and thoroughness it brings in comparison to the existing processes of review. Applying generative AI in ways that utilise "knowledge querying" and a form of "AI agent" for autonomous tasks (Teng et al., 2025) helps to move tasks from "messy" tasks to more consciously being part of a seamless system.

## 2. *Testing current generative AI: an experimental methodology*

To assess the viability of generative AI for internal peer review, we use an experimental methodology. We tooks papers presented for the Unit of Assessment 17 "Business and Management Studies" in REF 2021 to undertake a form of "reverse engineering" on the published percentages of scores for a selection of institutional submissions. The individual scores for outputs from REF 2021 are not available and never will be. "Once the sub-profiles were complete, the scores for individual outputs were no longer required and have been destroyed. In accordance with data protection principles, we no longer hold the scores for individual outputs as they constitute personal data, which should not be held for longer than required to fulfil their purpose" (2021.ref.ac.uk/faqs). As a consequence we have utilised the percentage allocations of scores of individual institutional returns to the target Unit of Assessment (UoA17) to determine if the scores the LLM ascribed to the individual papers will then calculate to something approximating the known combined percentages reported on an institution by institution basis. We have respected the REF determination that personal data is involved by anonymising the institutions and papers that we examined (e.g. "University A", "Paper 10"). The AI prompts and the software tools built to automate the research tasks are shared [available on request] so they can be applied to further collections of papers.

*2.1 Reverse Engineering REF2021*

A combination of institutions were used ranging from small to larger institutions. The AI assessment was only applied to journal outputs from the submissions. In total, 14 institutional returns for UoA17 were examined but only 11 were used in the final analysis as the remaining three had submitted multiple non-journal outputs or had too many inaccessible outputs that resulted in less than 90% of the submission being available. Smaller submissions are particularly affected by this combination of circumstances. The total sample was represented by 822 individual papers each reviewed five times with the mean of these reviews being used as the reported score. Working backwards from the reported total scoring percentages for each institutional submission we then identified how many papers would fall into each bin of the 4 point scale. We then produced AI-generated percentage scores for each of the papers to calculate the projected score boundaries for each institution. Where outputs were missing from an institutional set a cumulative distribution function was used to determine the potential range for the projected boundary score. A set of boundary points from across the entire sample was then calculated. These calculated scores were 49.35% dividing 1* and 2*, 58.52% between 2* and 3* and 69.06% between 3* and 4*,

The approach used does raise the potential for identifying any bias for, or against, an individual institution against the original process. To ensure robustness of the experiment, each paper was sampled at least five times against the LLM, with the mean of the results used in subsequent calculations and reporting. The tool that we developed for the analysis also provided for setting the temperature or variability in the AI-based responses. This was consistently set at 0.2 for all the papers that were reviewed and was used to mimic "human" variability in judgement and decision making. The temperature did make a difference in the maximum and minimum scores presented by the AI (Table 1) that reflects a bell curve of variation across the full set of samples reflecting the systematic nature of the system - lower temperature settings would move this variation down towards 0.

| Percentage of papers | Total Papers | Score variation (max/min) |
|---|---|---|
| 0.5% | 4 | 0 |
| 5.8% | 48 | 1 |
| 17.6% | 145 | 3 |
| 28.1% | 231 | 5 |
| 24.0% | 197 | 7 |
| 17.5% | 144 | 10 |
| 6.1% | 50 | 15 |
| 0.2% | 2 | 25 |

Table 1: Variations in score across the full set of papers examined (min 0/max 15.77)

To automate the process, and encourage reproducibility, two software tools were developed. The first was a web gatherer or robot [available on request] that ingests the REF results spreadsheets (from results2021.ref.ac.uk) and then attempts to collect the documents that are listed. The robot relied on the supplied DOI. If unpaywall.org offered an open access location it was used otherwise the DOI was resolved to the version of record. Often this pointed to a password protected page. Rather than circumvent acceptable use policies and publisher restrictions the discovered URLs were recorded and subsequently downloaded manually. The purpose of this robot was to gather as close to 100% of the available resources as automatically as possible. The second tool is presented as a web interface [available on request] that provides a consistent interface for generating the review scores. The interface accepts Word and PDF documents, automates sending the request five times to the large language model (LLM) which in this case was ChatGPT 4.1 (OpenAI *et al.*, 2023). This interface also provides format and error checking of the files. The interface managed the throttling of requests and retry requests if any failed. This software development was itself assisted with AI tools - specifically the Claude 3.5 Sonnet model - this support was invaluable as there is some complexity required to deal with the variability in file formats and interacting with the ChatGPT API. The full set of anonymised results including component scores and generated feedback comments are stored as an Excel file to simplify filtering and sorting [available on request].

*2.2 Iterative refinement of prompts*
To optimise the AI model's performance, we used an iterative process to refine the prompts used to generate the scores. By examining the model's outputs and identifying areas of inconsistency or inaccuracy, we adjusted the prompts to improve quality and stability in the generated scores and comments. This iterative refinement process is an inevitable aspect of research that utilises generative AI and requires acknowledgement. As a result of using the application programmers interface (API), and distinct from public chat interfaces, we provided two parts to the prompts being used: 'system' and 'user' prompts. The full text of the individual paper was appended to the user prompt. The increased context size - the size of the prompt that could be sent to the system - introduced with ChatGPT-4o and continued with 4.1 enables very long requests to be sent. The prompt also included a request for a short commentary of each criterion in the response. The additional information was used as a way of confirming that the paper was actually being parsed by the AI. In the initial testing through ChatGPT's public chat interface it was evident that occasionally the paper itself was not always included and the AI completely hallucinated a paper and its contents and then scored it in a very plausible way.

The user prompt asks the large language model to, "Critically score this paper on a 100% scale using the full range and high precision based solely on the three criteria of rigour, originality and significance. Rigour is the extent to which the paper demonstrates intellectual coherence and integrity, and adopts robust and appropriate concepts, analyses, sources, theories and/or methodologies. Originality is the extent to which the paper makes an important and innovative contribution to academic understanding and knowledge in the field. Significance is the extent to which the paper has influenced, or has the capacity to influence, knowledge and scholarly thought, or the development and understanding of policy and/or practice. Provide the response in this consistent format with the percentage score being contained within [] and the sections separated with the | character:  rigour:[score] short explanation | significance:[score] short explanation, | originality:[score] short explanation" and the system prompt defines the context in which the user is working, "You are a hypercritical reviewer of academic papers who scores out of 100% and uses the full range of scores with high precision. Work scored over 75% is considered world-leading quality and is less commonly seen, work over 50% is seen as internationally excellent but which falls short of the highest standards of excellence and is often seen, work over 25% is recognised internationally and is regularly seen and work under 25% is recognised nationally."

Through the process of iteration, the system prompt with the specific REF-based definitions was added specifying what constitutes the score of a paper to assist in giving consistency to the responses. The user prompt utilises the definitions for the three criterion used by REF to assess outputs. The phrasing was modified to place these definitions in the present tense in contrast to the formally documented versions. These modifications need to be acknowledged as they produced more consistent responses when tested repeatedly against the same paper. It is also significant to note that each request is sent to the AI as a new request meaning that it is self-contained and isolated from any other request. Each time the user and system prompts are used they are not part of a single ongoing conversation as might be the case when using the public interface chat. This avoids the potential that papers considered earlier in a chat session might influence the scores for subsequent papers. Setting the percentage barriers for the REF definitions of its scoring criteria in the system prompt is arbitrarily laid across the entire scoring range. However, drawing on the evidence from the experiment it is clear that while this does fully guide the scoring as the AI avoids using the full range seemingly regarding a score under 40% as some form of failure. This reflects deeper - and inaccessible - built-in system level instructions that appears to show that the LLM is always endeavouring to please with its responses (Wallace *et al.*, 2024). As a result an important iteration was to add the words "critical" and "hypercritical" to the pair of prompts. This reduces the most effusive of responses. The result is a prompt that tends to score within a range that is commonly used by UK academics for assessing student works with most scores falling between 50% and 70% and occasional dipping towards 40% or rising up to 80% (Yorke et al. 2000). Our final overall score boundaries of broadly 49/59/69 will be familiar boundaries to UK academics in the context of grade and classification boundaries. More speculatively, it is quite possible that the LLM does recognise this distribution pattern of scoring in relation to the rank ordering of academic work presented by students as a more general rule and is overlaying this pattern onto each request. To borrow the parlance of developers, "this is not a bug, it is a feature"

In the interests of reproducibility, the end result of these modifications and iterations are the prompts reported above and are the only prompts that have been used in the reported experiment. The code behind the web interface is also shared [available on request] so that it can be used to enable other researchers to access the same method without additional technical knowledge. It should be noted that in the shared Excel the 4-point-scale and 12-point-scale calculations were speculative. The current web version now reports these results based on the overall boundary findings of this paper.

*2.3 Acceptable margins of error and wider application*

Close attention has been paid to the potential for errors or even simply random numbers to be introduced into the results. The outputs of the generative AI have been accepted critically and with full acknowledgement that these are the results of using a statistical model that produces the most likely response to a prompt based on the training data of the Large Language Model (Cain, 2023). As an experiment, the range of scores that would be produced in response to the fixed prompts with the variable of the 822 different academic papers was completely unknown. As the purpose of the experiment was to explore the benefits of AI within the internal review process we turned to the REF2021 results as a potential benchmark to calibrate the AI scores against a partially known knowledge base. However, there is an imperative to find workable solutions for wider applications that will work for future REF exercises given that, "decoupling also appeared on the above list of the most burdensome changes… Several pointed to the burden of reviewing a large pool of outputs and having internal mock exercises to select outputs explaining the burden" (REF 2021- Cost Evaluation Final report).

## 3. AI reviewing results

The results from all of the AI reviews have been gathered into a single Excel file [FIGSHARE]. The file has been modified to remove the paper title, institution and filename to prevent casual personalised judgements. Access to de-anonymised reviews is possible in the interests of extending the science, supporting the reproducibility of this work and providing feedback. Each paper is presented as a single line in the CSV indicating the overall calculated mean score from the five samples as well as the highest and lowest scores. The individual scores for rigour, significance and originality are also shown along with the most critical comments generated by the AI - these are drawn from the lowest scores for each criteria. These comments do not identify authorship but do, occasionally, mention a specific case study, company or location.

The calculations for each institutional set of papers including applying the cumulative distribution function where not all papers could be obtained (or books and chapters were included in the institutional submission) indicates the potential range for a boundary between each point on the four point scale. The results are visually represented in the following section and are drawn from this available data.

As a further confirmation of the process, a post-analysis check of duplicate records - papers submitted independently by two institutions - was conducted. Only 36 outputs (4.4%) from the institutions were duplicates. Considering the absolute differences between these pairs and whether this difference produces the potential for different four-point scores when compared against the overall calculated boundaries. Of the 18 pairs found, fifteen of the results were consistent. Of the remaining three, two of the results produced a crucial inconsistency between the 2* and 3* boundary (Table 2). A result that suggests a 90% confidence rate in selecting outputs solely using AI.

|  | Paper 1 | Paper 2 | Absolute difference | Cross boundary | Nominal score |
|---|---|---|---|---|---|
| Pair 1 | I/113 | J/114 | 8.49 | No | 4* |
| Pair 2 | I/166 | K/66 | 12.4 | Yes | 3*/4* |
| Pair 3 | I/144 | K/104 | 0.47 | No | 4* |
| Pair 4 | G/10 | K/38 | 5.03 | Yes | 2*/3* |
| Pair 5 | D/33 | K/4 | 6.31 | Yes | 2*/3* |
| Pair 6 | I/81 | K/39 | 3.54 | No | 3* |
| Pair 7 | K/103 | M/82 | 2.54 | No | 4* |
| Pair 8 | I/157 | K/121 | 0.83 | No | 4* |
| Pair 9 | I/53 | K/37 | 0.14 | No | 3* |
| Pair 10 | I/139 | K/82 | 4.94 | No | 4* |
| Pair 11 | I/164 | K/116 | 1.75 | No | 4* |
| Pair 12 | K/110 | M/86 | 4.94 | No | 4* |
| Pair 13 | I/170 | K/127 | 1.43 | No | 4* |
| Pair 14 | I/172 | K/131 | 0.45 | No | 4* |
| Pair 15 | G/26 | K/52 | 0.07 | No | 3* |
| Pair 16 | I/86 | K/57 | 1.73 | No | 3* |

| Pair 17 | I/132 | K/95 | 0.01 | No | 4* |
| Pair 18 | D/94 | J/70 | 0.42 | No | 4* |

Table 2: Comparison of duplicate papers submitted by different institutions using overall calculated boundaries

### 3.1 AI scoring of outputs

The papers were reviewed by AI through the web interface. In selecting institutional submissions for analysis, purposive preference was given to those cases in which all, or the majority of, outputs were journal articles. This approach was adopted to enhance the completeness of the sample, as journal articles are more readily available. Consequently, institutions with more diverse submissions were typically excluded from the analysis. The outcomes could then be visualised using a dot plot of the papers showing the minimum, maximum and mean scores. In the example case of University B (Figure 1), a total of 44 outputs and 100% of this return were scored through AI. In the case of this particular return there were papers across the full range of scores in REF 2021.

The dot plot shows the range of results for an output that is the result of sending the same prompt to the large language model five times. For University B there is a mix of papers with extreme variations (2, 30, 35, 38) as well as papers with little or no variation (5, 6, 14, 15, 16). It is also notable that the two papers with the lowest and highest mean scores sit away from the rest of the pool. For internal evaluation, where detailed investigation of papers on classification boundaries is an important part of the process, this combination of evidence may help to indicate papers that may span boundaries and particularly between 2* and 3*. In this example, University B's papers 26, 30 and 36 would benefit from additional scrutiny. For the paper 26, human intervention may argue for its inclusion in a more discriminating submission pool while 30 and 36 might ultimately be discarded.

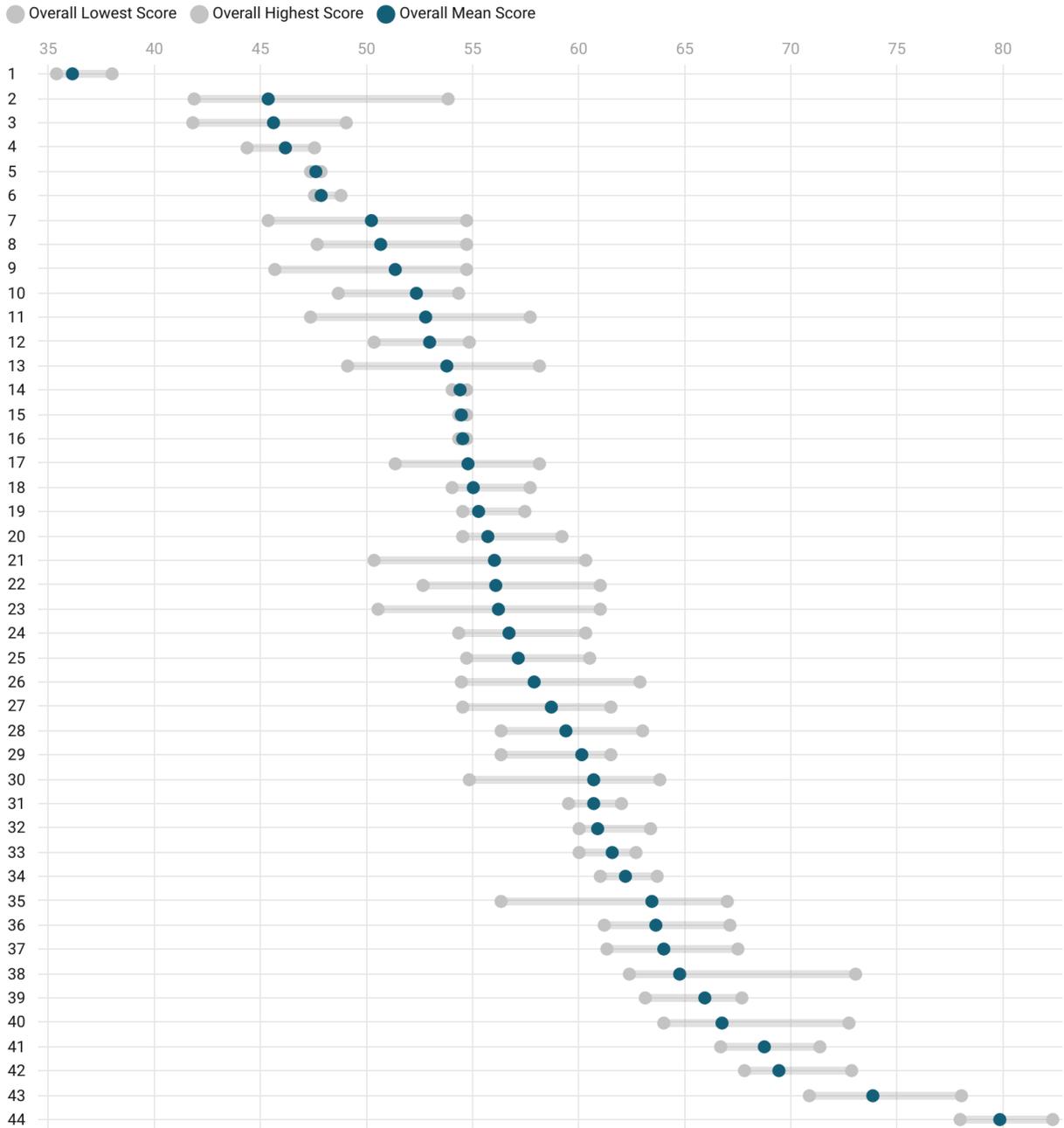

Figure 1: Individual outputs from University B - showing mean (dark) and min/max scores (grays) - calculated score boundaries are 45.88% (1*/2*), 55.75% (2*/3*) and 69.00% (3*/4*).

A further ten institutions were fully examined in the same way. For seven of these institutions where not all the outputs could be retrieved, the application of the cumulative distribution function defined the maximum and minimum boundary ranges for each score. A further three institutions where the combination of too few outputs and too many unavailable outputs resulted in them being discarded from subsequent calculations as these missing values resulted in the potential boundaries for scores extending across more than one scoring range (this excluded Universities C, F and L). The papers were then individually plotted (Figure 2) with the overall boundary range calculated from the fully analysed set of institutions. Papers that straddle boundaries are indicated with an 'X' - that is, papers that were calculated with a projected score based on the institution-based reverse engineering but have a different score based on the overall boundary calculation.

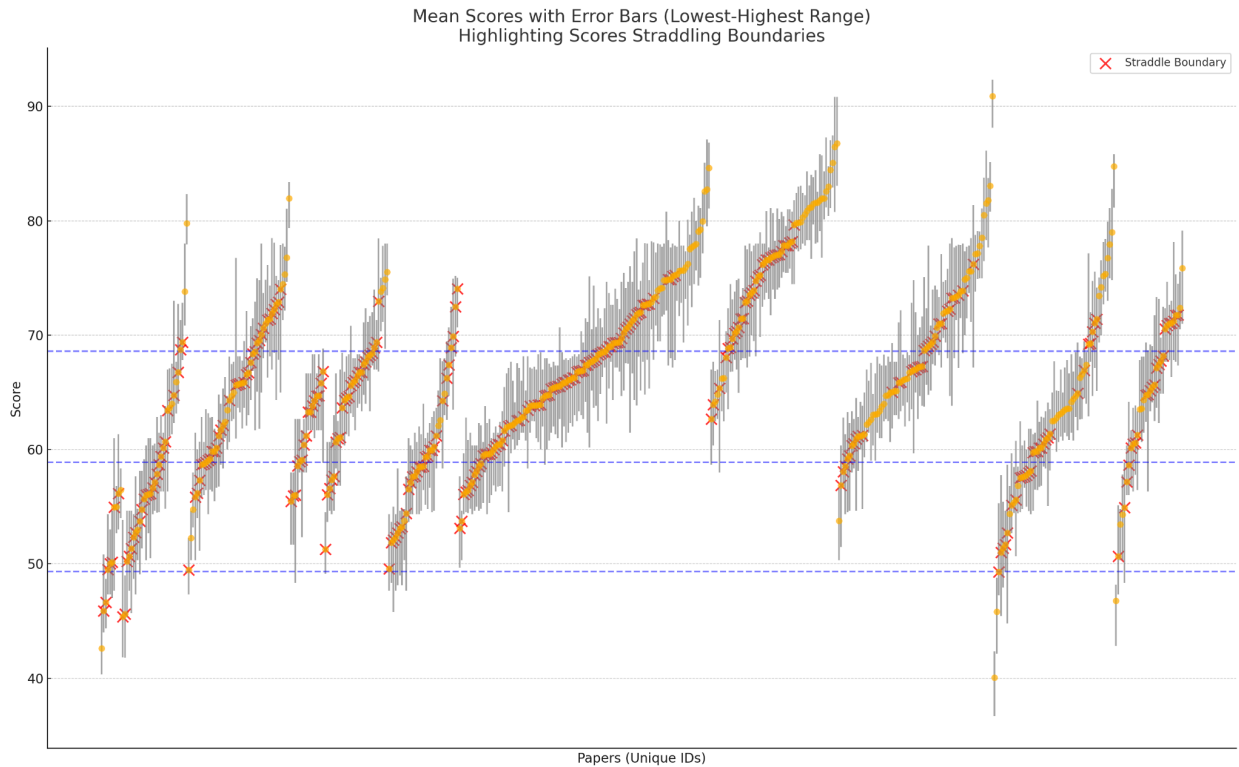

Figure 2: Complete plot of all scores with error bars and overall score boundaries

Across the individual institutions the calculated scoring boundaries ranged between 45.75% and 53.50% for 1*/2*, 55.75% and 63.00% for 2*/3* and 63.25% and 76.58% for 3*/4*. It is noticeable that these ranges are particularly distorted by two institutional results with University E exhibiting a very low boundary for 3*/4* outputs and University J showing very high boundaries for bother 2*/3* and 3*/4* (Figure 3). Bringing the results together from the institutions that produced a set of boundary scores the overall range boundaries were 49.35% between 1* and 2*, 58.52% between 2* and 3* and 69.06% between 3* and 4*.

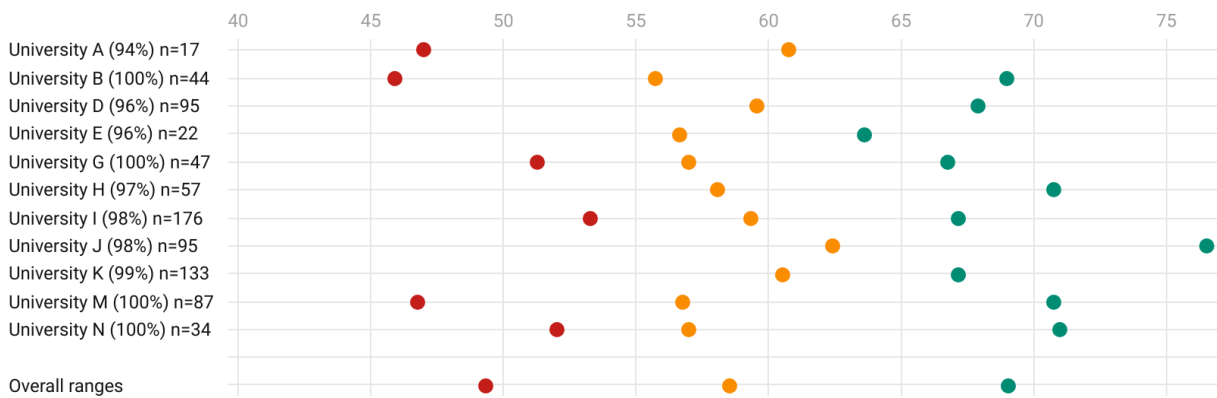

Figure 3: Summary boundary ranges calculated from individual AI scores based on the % of grades documented for each submission

## 4. Discussion and Findings

The implications for using generative AI to internally evaluate research outputs are far-reaching. By automating much of the process, institutions can free up significant time and resources, allowing academics to focus on their own research activities. The precision and consistency offered by an AI-assisted approach could also lead to fairer selection of outputs that does not embed unconscious bias of decision makers. However, the adoption of generative AI in this way does raise questions about trust and accountability. The academic community benefitting from this approach would also need to have confidence in the fairness, accountability and reliability of any AI-assisted system being used. Building this trust would require extensive validation and transparency in the development and deployment of the approach. As with many aspects of technology the initial investment of resources required to achieve this trust may exceed the current ongoing resources invested in existing - largely manual - processes.

Proposing an AI-assisted approach for any wider purpose beyond internal review requires consistent approaches and similar levels of trust and confidence to extend between institutions. The complexity of the relationship between large language models and the language found in reviewed outputs emphasises the need for transparently shared prompts, temperature settings and agreed common models to ensure comparability between any scores that are generated. This challenge extends beyond the simple technical possibilities of AI-assisted scoring and directly faces the organisational research culture of UK higher education and its collective preparedness for innovation or change.

*4.1 The value of AI-assistance for internal evaluations*

The results of the experiment offer some cautious opportunities. From a systems theory perspective, the use of AI in this context constitutes a cybernetic substitution. In other words, the offloading of internal review tasks from overextended subsystems to maintain institutional coherence. This has far wider reaching implications for the use of AI-assistance in other aspects of any organisational activity that requires large volumes of repeated assessment of highly variable but structured qualitative material including for example, court judgements and the personal development records of employees. For the practical purposes of internally evaluating academic works, the AI-generated scores do highlight some edge cases that exhibit large variations in their scores across the five samples or between the three criteria for evaluation. Papers that exhibit these attributes may, in the right context, be regarded as groundbreaking, innovative or, at least, unusual. For requirements that require risk aversion, including the preparation of a pool of work for external assessment, this same approach may provide a means to flag up papers that could produce unpredictable results. For papers that are uncontentious and sit in the middle of a predicted grade boundary the AI-assistance may help to confirm the evidence of other forms of external metrics as well as that of human reviewers. The "safest" outputs of this type scored at 3* and 4* would represent the core pool of a submission for external evaluation. Applying this same form of analysis to an individual's own output trajectory may also provide feedback and guidance for the pathway they are endeavouring to chart by revealing a balance of activities between risk, advancing core knowledge and internally coherent quality.

*4.2 Policy implications for research assessment*

The successful assistance of generative AI in internal peer review could also have significant implications for the REF process itself. If LLMs can consistently produce accurate and precise assessments of research quality, there is an argument for directly incorporating this technology into national research assessment exercises. By using a single, consistent AI model for evaluating all research outputs, external assessments of quality could potentially eliminate the variability introduced by human reviewers and streamline the entire process.

Moreover, the improved precision offered by generative AI could enable a more granular approach to research assessment. Instead of relying on broad quality categories (i.e. a four-point scale), research evaluations could adopt a continuous scale, allowing for finer distinctions between research outputs. Increased precision could, in turn, lead to a more nuanced and targeted distribution of research funding, as the system could identify and reward high-quality research with greater accuracy. AI-assisted scoring also makes it possible, with specific prompts, to identify and reward research that show indications of breakthrough thinking, innovation or new knowledge.

However, the adoption of generative AI in national assessment does require careful consideration and planning. The development and validation of the chosen AI model would need to be transparent and subject to rigorous testing to ensure its fairness, reliability, and robustness. Additionally, the assessment process would need to establish clear guidelines and protocols for the use of AI in research assessment. Beyond respecting the principles of DORA there is a need to address issues such as data privacy, intellectual property rights and the underlying potential for algorithmic bias.

*4.3 Theoretical Framing: Viable System Model and AI-Enabled Substitution*

The experimental methodology presented here can further be productively theorised through the lens of Beer's Viable System Model (VSM) (Beer, 1981). VSM is described as a cybernetic framework (and is a precursor to further work that has come to be labelled as artificial intelligence) designed to describe the necessary functions that must exist within any viable, self-regulating and adaptive organisation. The model identifies five key system functions (System 1–5) that enable autonomy at the operational level while preserving coordination, control, intelligence, and identity at higher systemic levels.

In the context of internal research evaluation for the REF, the subsystem responsible for internal review and quality assurance aligns with System 3 — the locus of control and resource optimisation. This subsystem is increasingly overburdened by the bureaucratic and labour requirements of evidencing quality across hundreds of outputs, leading to a systemic inefficiency that ultimately compromises the viability of the entire organisation (Beer, 1981). This is an undesirable situation and one that all organisations seek to avoid.

Introducing GenAI based assistance into the internal evaluation process constitutes a functional substitution within System 3. Rather than advocating for the elimination of human judgement found in System 2 (coordination and stability) or in the strategic policy-setting of System 5 (responsible for identity and governance), GenAI supplements operational oversight of System 3 by supporting a high degree of automation within repeatable evaluation tasks. This substitution maintains the overall viability of the higher education institution by absorbing these key aspects of complexity, helping to reduce systemic entropy that is articulated through unmanageable workload, and enabling other systems to operate more reflexively and efficiently.

What is being described - and advocated for - here does not represent a replacement of systemic functions but a redistribution of the associated workload. An observation that also aligns with the recursive nature of the VSM in that systems are made of other small systems. Subsystems within this institutional system, such as individual Units of Assessment (UoAs), all need to operate as viable systems in their own right. When GenAI is deployed at this level to provide preliminary scoring, it becomes an adjunct to specific and localised System 3 functions. This arrangement can then enable better prioritisation of the scarce resource of human-based evaluations and attention.

Reallocation preserves organisational coherence of the VSM through System 4 (planning and intelligence) by providing fast - and comparatively faster - feedback loops that inform future output trajectories and strategic alignment. From a VSM perspective, AI assistance is not reduced to the role of being a cost-saving tool. AI assistance is an exemplar of how a "cybernetic" enabler can absorb operational variety and even restore stability and equilibrium across an organisational system that is under pressure from external accountability demands.

## 5. Next Steps, contribution and conclusion

The evidence presented in this paper shows that the use of generative AI for internal evaluation of outputs is a promising approach to improve the efficiency, consistency, and precision of the internal evaluation of outputs. The methodology presented here provides a framework and the tools to deploy a viable approach within an institution for, at least, business and management studies focused works.

As a result, this work illustrates how generative AI can become an active component in viability maintenance by absorbing operational variety and automating a particularly resource-intensive process in (UK) academia. While Beer's VSM presupposed human-mediated decision-making across all system layers, this paper empirically shows that LLMs can be structurally integrated into System 3 functions and as a consequence redefining the boundary between human and machine agency within viable systems. This extends VSM theory by demonstrating the integration of (generative) artificial intelligence into System 3 functions, showing that machine agents can perform viability-preserving substitutions in highly institutionalised systems without producing any degradation in systemic integrity.

However, the results do raise some questions and do present some limitations. The AI-assisted scoring does not take the publication year into account when it makes judgements about, for example, originality. A paper written and submitted for REF 2021 could have been written up to a decade previously with data gathered even earlier. The language model is trained to a point in time that is much more recent. It is possible that something that was original in 2015 is not the case in 2025. There is also a counterpoint to this statement: how do human reviewers currently account for this same shift over any given REF period?

Despite obtaining positive results with the experimental method presented, the unknown underlying workings of the large language model do present challenges. There is some indication that results are clustering around some form of internal system boundaries. This can be seen in the detailed results from University B (Figure 1) where the mean result appears to drift towards the 55% and 65% mark to the extent that the mean drifts towards the minimum or maximum values if they are closer to these percentage points. The lower standard deviation of the 2*/3* boundary (of 2.02) does suggest that the model is already familiar with a lot of academic output and is potentially recognising that papers presented for external evaluation are of better quality than the bulk of those that are never submitted for assessment. The implication being that the ChatGPT large language model has already been trained in some way on REF papers as examples of quality academic work.

The range of unknowns in commercial AI tools and the lack of opportunity to train an LLM specifically on the complete REF2021 data (Bisson, 2024) requires an approach that can present statistically sound scoring boundaries. A repeatable methodology can then be applied to all outputs under examination through future versions of LLMs that will almost certainly improve the statistical confidence of the output scores. A statistically consistent approach also mitigates the concerns of some researchers, "Using ChatGPTs for academic evaluations can provide consistent feedback, but it may not always be entirely accurate, relevant, or useful. There are challenges in ensuring the objectivity and quality of comments. Additionally, the increased use of ChatGPTs in evaluations may perpetuate biases related to race, region, or class" (Mehta *et al.*, 2024). Ultimately the goal of using AI assistance for internal evaluation of outputs would be to support a process that is better than the one currently in use. But this could equally be applied to national assessment processes too. AI when used to maximum effect should extend capability and confidence for an activity not merely substitute existing human actions. "Using AI tools may help … not merely to save reviewers' time, but also to uncover biases in decision-making. Uncovering such biases may help to develop approaches to reducing or eliminating their impact" (Checco *et al.*, 2021). A systematic review of the biases that may exist between REF Units of Assessment, individual institutions, or temporally when considering the originality of a paper are all potentially productive avenues for further research utilising the same or similar methodology presented here.

A final limitation is a very human one. Given that human reviewers are subject to many forms of influence and unconscious bias an experimental methodology that uses the scores (and implied score boundaries) from the previous REF is also founded on data with this same bias. It is possible that the variations shown in the score boundaries between institutions may themselves quantify the extent of this bias across the sample and that the overall score boundaries that we ultimately present are unconsciously representing the starting boundaries that reviewers utilise in their own scoring. A somewhat circular argument that, if it could be proven, would also further strengthen the case for AI-assisted reviewing more widely.

With these cautions and speculation, the presented experiment does reveal that LLM-based automation can serve as a recursive independent but reusable subsystem within recursive systems (in other words, research Units of Assessment within a university, a university within a higher education sector), illustrating a practical application of modular viability - and of systems modularity. What is described here can be applied within other subsystems with little or no modification to provide a shared consistency and comparability. This makes a contribution to Beer's emphasis on the importance of systems recursion with a contemporary example in which the AI assistance functions as an internal regulator within a nested viable system that does not reduce the overall system's autonomy or responsiveness.

There are many opportunities for further research and wider application of the methods and tools presented here that go beyond the direct use of supporting the preparation of a submission pool of outputs for external evaluation. Providing a service to support authors prior to submission would help overcome the reticence experienced by some early career researchers to share their work for feedback. More broadly, systematic and bibliometric approaches to analysing and interpreting published research are an inherent aspect of large language models. This built-in capability presents an opportunity for new and innovative methods that better reveal the purpose of the research and should lead to more impactful and useful research outcomes.

As theorised through the Viable System Model, the role of AI in institutional context should be to preserve the organisation's capacity to function viably amid escalating external demands rather than, more mundanely, leading to the displacement of human judgment. However, Roberts and

Candi (2024) argue that generative AI role is seen as a means towards enhancing employees' work and to make employees' jobs "more fulfilling". The substitution of routine labour on irregular and variable data with underlying algorithmic consistency helps maintain institutional adaptability, responsiveness and enables, for higher education institutions at least, to preserve focus on knowledge generation. An observation that reinforces Just's (2024) advocacy for innovation stakeholders to "rethink current roles and responsibilities in AI-based innovation processes" and reinforces a central tenet of VSM thinking that adjusting internal structures and role distribution is a pivotal remedy to systemic dysfunction that strengthens system efficiency.

As academia deals with the challenges of limited resources and increasing demands for accountability, AI-assisted solutions do offer a useful step forward, enabling institutions to focus on their core mission of advancing knowledge and sharing these findings to the widest possible audiences.